\documentclass[12pt]{iopart}

\usepackage{epsfig}
\usepackage{graphicx} 
\usepackage{dcolumn} 
\usepackage{bm} 
\usepackage{iopams}

\newcommand{\ex}[1]{\times 10^{#1}}
\newcommand{\npe}{n_{\rm e}}
\newcommand{\nb}{n_{\rm b}}
\newcommand{\dv}{\Delta v}
\newcommand{\vb}{v_{\rm b}}
\newcommand{\vph}{v_{\phi}}
\newcommand{\nr}{\nb / \npe}
\newcommand{\wpe}{\omega_{\rm pe}}
\newcommand{\wrr}{\omega_{\rm r}}
\newcommand{\wi}{\omega_{\rm i}}
\newcommand{\ko}{k_{0}}
\newcommand{\phb}{\phi_{\rm b}}

\newcommand{\Wi}{\wi / \wpe}

\newcommand{\T}{\wpe t}
\newcommand{\Z}{\ko z}

\newcommand{\fb}{f_{\rm b}}
\newcommand{\wt}{\omega_{\rm t}}
\newcommand{\V}{v / \vb}

\newcommand{\Ph}{\phi / \phb}
\newcommand{\Wt}{\wt / \wpe}
\newcommand{\Dv}{\dv / \vb}

\DeclareFontFamily{OML}{eur}{\skewchar\font127}
\DeclareFontShape{OML}{eur}{m}{n}{<5> <6> <7> <8> <9> gen * eurm <10> <10.95>
  <12> <14.4> <17.28> <20.74> <24.88> eurm10}{}
\DeclareSymbolFont{greek}{OML}{eur}{m}{n}
\DeclareMathSymbol{\micro}{\mathord}{greek}{"16}

\begin{document}

\title[Preprint]{Phase-Space Holes in an Electron-Beam-Plasma}

\author{TAKEDA Tsuyoshi\dag \footnote[3]{Email: sttaked@ipc.shizuoka.ac.jp}
and YAMAGIWA Keiichiro\ddag}
\address{\dag\ Radiochemistry Research Laboratory, Shizuoka University, Ohya 836, Shizuoka, 422-8529, JAPAN}
\address{\ddag\ Department of Physics, Shizuoka University, Ohya 836, Shizuoka, 422-8529, JAPAN}

\begin{abstract}
It is shown in an electron-beam-plasma system that phase-space holes evolve dynamically in electron time scales from appearance to collapse. The holes are synchronized with a wave packet dominated by a beam mode, and their velocity radii depend on the packet crest amplitudes. This suggests that the holes are induced by a self-trapping phenomenon. It seems from the observed images that the trapped-beam is hard to be detrapped while the packet grows linearly, but easy to be done after that.
\end{abstract}

\pacs{52.35.Py, 52.35.Qz, 52.35.Fp}
\maketitle

\section{Introduction}
In a cold electron-beam-plasma system, it is well-known that a wave packet develops linearly in the first regime, obeying electron-beam mode properties described by a linear equation~\cite{Briggs}. But the packet can not continue to grow infinitely, so that saturation and damping occur in the next regime. For the reason, it is thought that such a wave packet is easy to be accompanied with nonlinear phenomena~\cite{Hasegawa,Goldman}, such as modulational instability~\cite{Intrator}, electron-beam trapping~\cite{Mizuno,Gentle} and so on. Yamagiwa {\it et al}~\cite{Yamagiwa} experimentally observed that a wave packet in saturation process evolves into a wave packet train as continuously emitting some wave packets. They argued with the nonlinear theory of Yajima {\it et al}~\cite{Yajima}, but could not make the evolution mechanisms clear. Meanwhile, from viewpoints of computer simulation results~\cite{O'Neil,Akimoto}, it has been interesting that electron-beam holes are present in phase-space as resulting from particle trapping due to a wave packet. In previous work~\cite{Takeda}, we attributed an abnormal spread below a phase velocity to electron-beam trapping as approach to the above mechanisms, though we could not observe any holes.

Since the holes evolve in electron time scales as well as the packet, it had been very hard for those evolutions to be explored in laboratory experiments. We use an energy analyzer~\cite{Stenzel}, and separate the analyzer signal into low and high frequency bands, and can capture phase-space holes consisting of beam-electrons in a cold plasma. In this paper, we clarify the dynamical evolutions of the holes, and discuss their behaviors correlated with the packet.

\section{Experimental System}

\begin{figure}[hbt]
\begin{center}
\includegraphics[width=8cm]{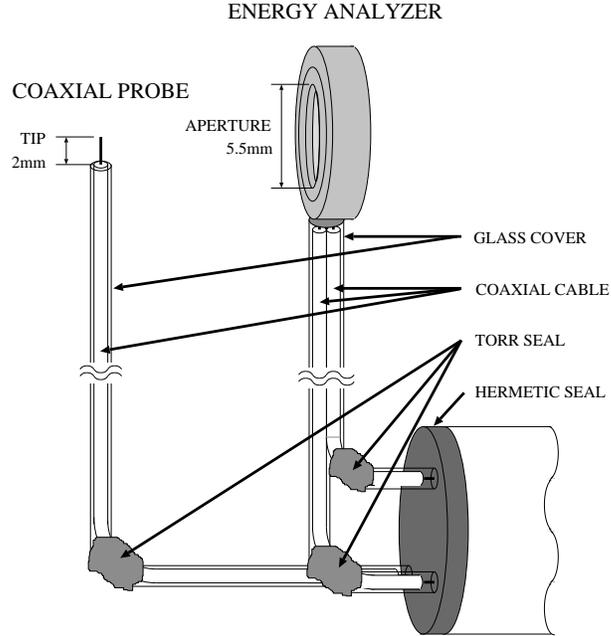}
\caption{%
A coaxial probe and an energy analyzer. The coaxial probe has a tip of diameter $0.3$ mm and length $2.0$ mm, which detects potential perturbations. The analyzer has an aperture of diameter $5.5$ mm, which consists of a discriminator and a collector shielded against electric fields. Discriminated electron currents derive a phase-space electron distribution.
}
\label{fig:probes}
\end{center}
\end{figure}

A cylinder chamber made of stainless steel, whose sizes are $0.26$ m in diameter and $1.2$ m in length, is filled with argon gas of low pressure $2.8\ex{-5}$ Torr. A cold plasma is produced by DC-discharging in the gas between four heated filaments and the chamber wall. Then the plasma is confined by full-line cusps~\cite{Leung} produced by twelve line-magnets mounted on the external surface of the chamber. An electron-beam gun at $z=0$ is mounted on an end flange of the chamber, and can emit a pulsed cold electron-beam with the diameter $50$ mm, the duration time $3.5\,\micro$s, and the mean energy of $\phb=50$ eV. The beam injected into the plasma behaves one-dimensionally along axial DC-magnetic field $0.01$ T induced by six external coils. A wave packet excited in this system is observed as potential perturbations by using a coaxial probe, whose tip is $0.3$ mm in diameter and $2.0$ mm in length. An energy analyzer~\cite{Stenzel}, which has the aperture of diameter $5.5$ mm, is adopted so as to observe a phase-space distribution of the beam. A collector in the analyzer is shielded against electric fields and can detect the beam currents discriminated by a biased grid facing the plasma. These probes are illustrated in Fig~\ref{fig:probes}.

These observations are synchronized with a test wave signal, which consists of the carrier frequency $90$ MHz and the envelop time width $50$ ns by the full width at half maximum (FWHM). The test wave signal is applied to a control grid of the gun at $t=0$ in order to excite a wave packet, and simultaneously triggers two digitizing oscilloscopes with $1\ex{9}$ samples per second. The packet signal detected by the coaxial probe is amplified by a high frequency amplifier ($0.1$-$1300$ MHz). The beam current signal detected by the analyzer is divided into two, and they are in parallel amplified by a low frequency (LF) amplifier of DC-$8$ MHz band and a high frequency (HF) amplifier of $8$-$1300$ MHz band. These amplified signals are individually received with time-averaging on two channels of the oscilloscopes, and those data are stored in PC. Ends of coaxial cables are all connected to matching resistances $50\,\Omega$. These detections are carried out at each of $128$ axial positions from $\Z=22$ to $75$ ($\ko$ defined below). The LF and HF signals of the beam current are synthesized on PC in each cell of the discriminating energy and the position. Eventually, the first derivatives of the synthesized beam current with respect to the discriminating energy at the positions give the phase-space distribution of the beam. Table~\ref{tab:para} lists typical parameters of the experiment. Here $\ko=\wpe / \vb$ is defined for convenience.

\begin{table}[hbt]
\begin{center}
\caption{%
Typical experimental parameters.
}
\label{tab:para}
\vspace*{0.20in}
\begin{tabular}{ll}
\hline
Plasma-electron temperature & $\simeq 0.8\,{\rm eV}$ \\
Plasma-electron density & $\npe \simeq 1\ex{14}\,{\rm m}^{-3}$ \\
Electron-plasma frequency & $\wpe / 2\pi \simeq 90\,{\rm MHz}$ \\
Beam-electron density & $\nr \simeq 0.3\,\%$ \\
Electron-beam velocity & $\vb \simeq 4.2\ex{6}\,{\rm m/s}$ \\
Initial electron-beam spread & $\simeq 0.05 \vb\,{\rm (FWHM)}$ \\
Special wave number & $\ko \simeq 135\,{\rm rad/m}$ \\
\hline
\end{tabular}
\end{center}
\end{table}

\section{Results and Discussion}

\begin{figure*}[hbt]
\begin{center}
\includegraphics[width=14cm]{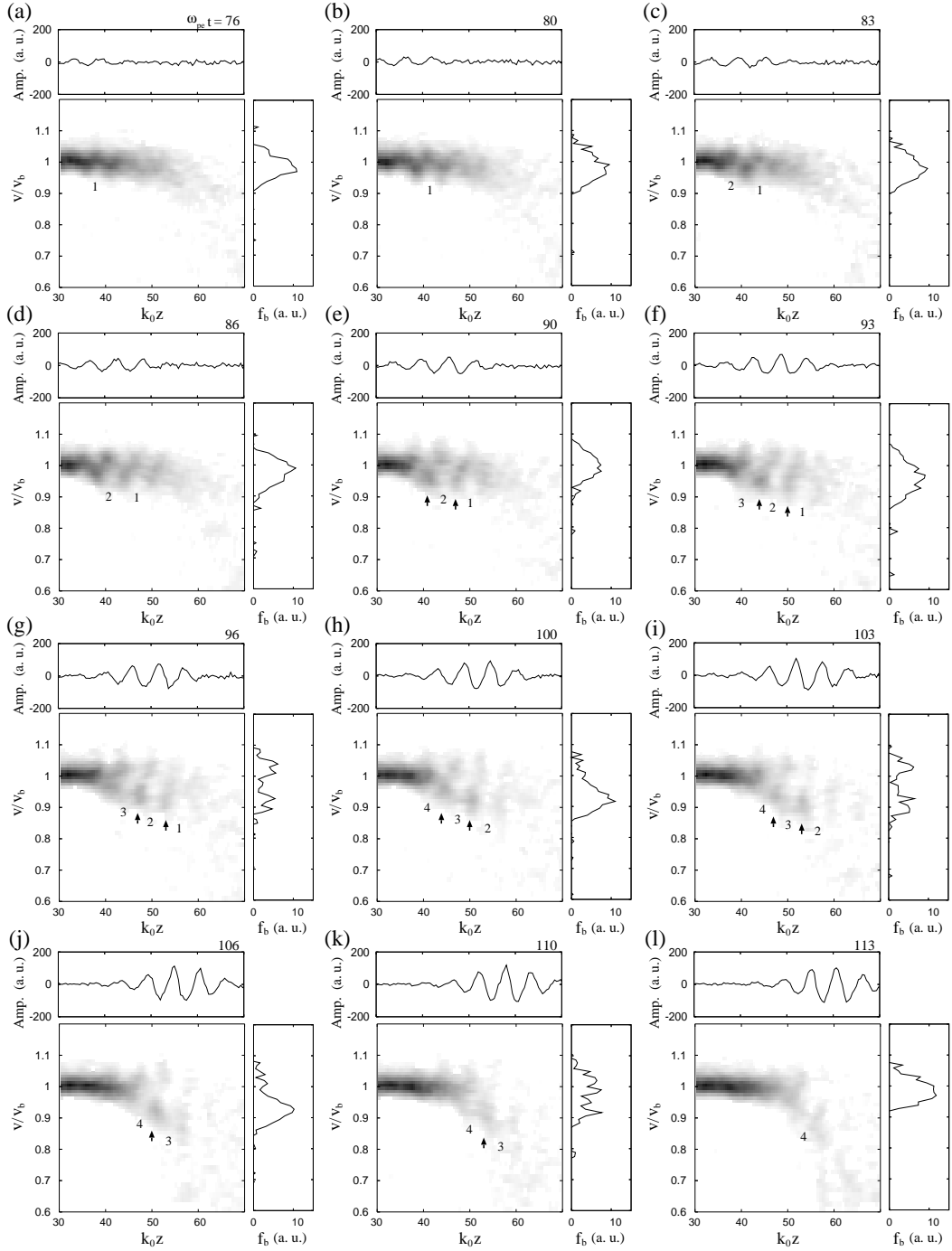}
\caption{%
A wave packet and phase-space holes in a weak electron-beam distribution $\fb$ for a period from $\T=76$ (a) to $113$ (l). Each upper left shows the packet amplitude vs $\Z$, each lower left $\fb$ vs $\V$ and $\Z$, and each right typical cross-section of $\fb$ at $\Z=52$. Dark contrast corresponds to the beam density in phase-space. Arrows point the beam bunching. Electron-beam holes are always synchronous with the packet crests.
}
\label{fig:holes}
\end{center}
\end{figure*}

Figure~\ref{fig:holes} shows a wave packet (upper lefts) and a phase-space distribution $\fb$ of the beam (lower lefts) in time intervals of $3.4 / \wpe$. Here dark contrast corresponds to the beam density in phase-space. Each right is typical cross-section view of $\fb$ at $\Z=52$. Here $v$, $z$, and $t$ are the velocity, the position, and the time, respectively. The packet exhibits the temporal evolution from linear growth ($\T<103$) to saturation ($\T \gtrsim 103$) processes. Four electron-beam holes emerge discretely as corresponding to positions of the packet crest. Since the holes evolve temporally, we assign a number for each from the downstream side so as to identify them easily. Note that random spontaneous waves excited by the beam, which are invisible in the averaging observation due to no synchronization with the test wave, scatter the beam especially in the downstream side, but the scattered beam can be treated as almost steady-state in this observation.

The typical profile of the packet is as follows: phase velocity is $\vph \simeq 0.92 \vb$, wave number $k \simeq 1.05 \ko$, frequency $\wrr \simeq 0.97 \wpe$, temporal growth rate $\wi \simeq 0.062 \wpe$, and group velocity in the linear growth process about $0.90 \vb$. The profile suggests that the packet in the linear growth process obeys an electron-beam mode in the two-stream instability~\cite{Briggs}.

Initially, as the packet is propagated downward, its amplitude grows linearly and the holes generate around $v= \vph$. For instance, the second and third hole, which emerge around $\T =83$, $93$, respectively, are gradually shaping fine circles with the packet growth, and as seen in the cross-section views at $\T =96$, $103$, their velocity radii become large with the packet amplitudes. The holes are in phase with the packet crests, and the radii are strongly correlated with the crest amplitudes. These results obviously prove that the beam is trapped in the potential wells of the packet, and the trapping starts from the initial stage of the growth process. The holes can be theoretically regarded as right-handed vortices~\cite{Hasegawa}, though their rotations are invisible. It is seen that the beam bunching (arrows), which is clear in the cross-section views at $\T =100$, $106$, arises between the adjacent holes. The bunching formation indicates that parts of the trapped beam are localized where electric fields are almost zero. When the packet shifts to the saturation process, the hole shapes gradually become unclear in the downstream side and appear to collapse, leaving their traces in the lower velocity side. Since the packet saturation is attributed to energy supply from the packet to the beam, it is likely to be detrapped partly in the downstream side and thus the hole collapsing seems to occur.

\begin{figure}[hbt]
\begin{center}
\includegraphics[width=10cm]{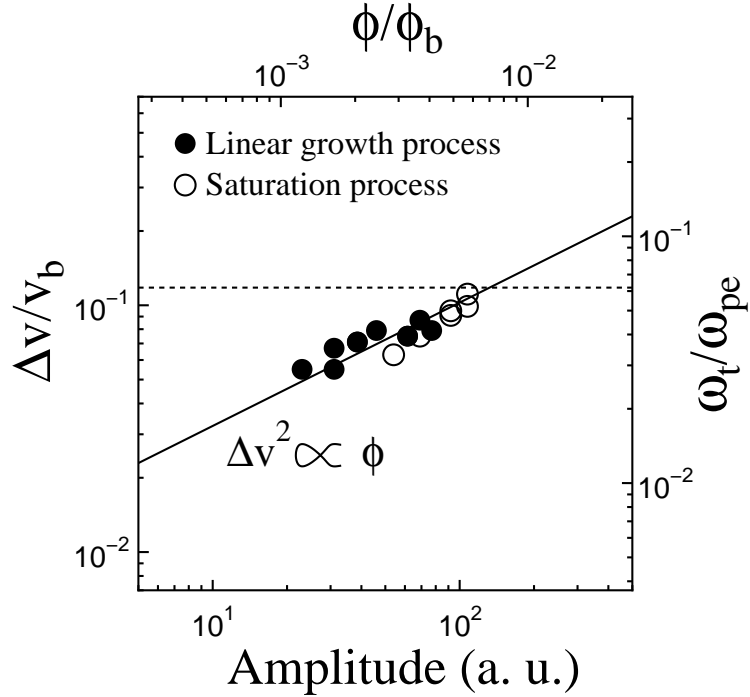}
\caption{%
Dependences of the hole velocity radii $\Dv$ on the packet crest amplitudes in linear growth (dots) and saturation (circles) processes. Upper horizontal and right vertical axes are the potential amplitude $\Ph$ and the bounce frequency $\Wt$, respectively. A broken line corresponds to the temporal growth rate $\Wi$ in the growth process. Plots agree with a solid line of $\dv ^{2} \propto \phi$.
}
\label{fig:depen}
\end{center}
\end{figure}

An electron trapped in a wave potential well bounces with the frequency of $\wt =(ek^{2} \phi / m)^{1/2}$ and rotates in phase-space with the velocity radius of $\dv =2 \wt / k$~\cite{Hasegawa,Akimoto}, which mean that $\dv^{2}$ is dependently proportional to $\phi$. Here $\phi$ is the amplitude of the wave potential, $e$ the electron charge, and $m$ the electron mass. Figure~\ref{fig:depen} shows the dependences of the hole velocity radii $\Dv$ on the packet crest amplitudes in the growth (dots) process and the saturation (circles) process, obtained from Fig.~\ref{fig:holes}. Axial ranges of $\Ph$ and $\Wt$ are calculated from $\Dv =0.023$ to $0.23$. A solid line is $\dv ^{2} \propto \phi$. A broken line corresponds to $\Wi$ in the growth process. We estimate that the maximum potential amplitude is about $6.0\ex{-3} \phb$, and then $\wt$ is about $5.8\ex{-2} \wpe$. Plots almost agrees with the solid line, which supports that the hole formation in Fig.~\ref{fig:holes} is due to the trapping effect.

On the other hand, electron detrapping is also regarded in that trapped electrons take away a part of wave energy. In the growth process, all plots of $\wt$ are less than $\wi$ as seen in Fig.~\ref{fig:depen}, in other words, the bounce times are longer than the growth one. Therefore this indicates that it is difficult for the beam to be detrapped. However the beam can be partly detrapped in the saturation process because $\wi$ eventually comes to be much less than all $\wt$. This result is consistent with that of Fig.~\ref{fig:holes}. The trapping suppresses the packet growth, but its effect seems so weak that the packet shape can not be deformed~\cite{Takeda}. Unfortunately, the packet train~\cite{Yamagiwa} is not observed in this experiment, but we assume that if the beam is strongly accelerated in the detrapping process, the second packet will be excited and then it be formed.

\section{Conclusion}
It is experimentally observed in an electron-beam-plasma system that phase-space holes consisting of beam-electrons emerge and develop dynamically. The holes are alive while a wave packet grows linearly, obeying an electron-beam mode in the two-stream instability. The development of the holes is correlated with that of a wave packet, and then the square velocity radii of the holes are almost proportional to the amplitudes of the packet crests. These prove that the holes are induced by a self-trapping effect. The packet growth is gradually suppressed by the trapping influence, so that the packet amplitude saturates eventually. The magnitude of the bounce frequencies, estimated by the velocity radii, implies that the beam is difficult to be detrapped in the growth process, but it is easy to be partly done in the saturation process. The holes appear to finish their lives as collapsing by the beam detrapping.

\ack
The authors would like to acknowledge useful discussions with Professor SAEKI Koichi and Associate Professor AKIMOTO Kazuhiro.

\section*{References}




\end{document}